\documentclass[12pt, preprint]{aastex}
\usepackage{amssymb}
\slugcomment{Submitted to ApJ}
\shortauthors{Ardila, Basri}
\shorttitle{GHRS observations}
\begin{document}

\newcommand \kms{km s$^{-1}$}
\newcommand \funit{$\rm{\ ergs\ sec^{-1}\ cm^{-2}\ \AA}$}
\newcommand \vsini{$v$sin$i$}
\newcommand \caii{\ion{Ca}{2}}
\newcommand \hei{\ion{He}{1}}
\newcommand \heii{\ion{He}{2}}
\newcommand \feii{\ion{Fe}{2}}
\newcommand \hal{H$\alpha$}
\newcommand \hbeta{H$\beta$}
\newcommand \rsun{R_\odot}
\newcommand \msun{M_\odot}
\newcommand \lsun{L_\odot}
\newcommand \degs{$^\circ$}
\newcommand \htwo{H$_2$}
\newcommand \lya{H$_{Ly\alpha}$}
\newcommand \mgii{\ion{Mg}{2}}
\newcommand \sii{\ion{Si}{1}}
\newcommand \siiv{\ion{Si}{4}}
\newcommand \civ{\ion{C}{4}}
\newcommand \nad{NaD}
\newcommand \siiii{\ion{Si}{3}]}
\newcommand \oiv{\ion{O}{4}]}
\newcommand \oi{\ion{O}{1}]}
\newcommand \cii{\ion{C}{2}}
\newcommand \ciii{\ion{C}{3}]}

\title{Observations of T-Tauri Stars using HST-GHRS: II. Optical and Near UV lines}

\author{David R. Ardila\altaffilmark{1}, Gibor Basri\altaffilmark{2}, Frederick M. Walter\altaffilmark{3}, Jeff A. Valenti\altaffilmark{4}, Christopher M. Johns-Krull\altaffilmark{5} }
\altaffiltext{1}{Astronomy Dept., Univ. of California, Berkeley, CA 94720, ardila@garavito.berkeley.edu}
\altaffiltext{2}{Astronomy Dept., Univ. of California, Berkeley, CA 94720, basri@soleil.berkeley.edu}
\altaffiltext{3}{Department of Physics and Astronomy, SUNY, Stony Brook NY 11794-3800, fwalter@astro.sunysb.edu}
\altaffiltext{4}{Space Telescope Science Institute, Baltimore, MD 21218, valenti@stsci.edu}
\altaffiltext{5}{Space Sciences Laboratory, Berkeley, CA 94720, cmj@ssl.berkeley.edu}

\begin{abstract}

We have analyzed GHRS data of eight Classical T Tauri stars (CTTSs) and one Weak T Tauri star (WTTS). The GHRS data consist of an spectral range 40 \AA\ wide centered on 2800 \AA. For 4 of the CTTS we have nearly simultaneous optical observations which contain \hal, \hbeta, \hei, \nad, and the \caii\ infrared triplet. The \mgii\ resonance doublet is the strongest feature in the 2800 \AA\ range. This line has a fairly wide and symmetric emission component ($\sim200$ to $\sim300$ \kms\ for the CTTSs), with a narrow central absorption and a wide blueshifted absorption superimposed to it. The narrow central absorption width and equivalent width are inconsistent with being due only to ISM clouds described in the literature, which lead us to conclude that it is partially due to non-LTE processes in the emission line region itself. The emission profile closely follows \hal. Its large width in CTTS cannot be due to the Stark effect and we suggest that it is due to supersonic turbulence. All the stars show blueshifted absorptions that are evidence of outflows (terminal velocities $\sim300$ \kms), with multiple flows observed in two stars. We show evidence that the wind is not spherical, with wind signatures being stronger for lower inclinations at a given accretion rate. We briefly compare other optical lines with the hot transition region lines observed in CTTS.
 
\end{abstract}


\section{Introduction}

In this paper we continue the analysis of Goddard High Resolution Spectrograph (GHRS) spectra for pre-main sequence stars. The first part of this analysis has been reported in \citet{ardila2001} (Paper I). Our sample consists of 8 Classical T Tauri Stars (CTTSs) and one Weak T Tauri Star (WTTS). Here we will concentrate in the \mgii\ resonance doublet at 2800 \AA\ and its relationship to (simultaneous) optical observations.

The \mgii\ resonance doublet (2795.53 and 2802.71 \AA, the k and h lines, respectively) is the strongest near-ultraviolet feature in the spectra of T Tauri stars (TTSs). Analyses of high resolution IUE observations of \mgii\ in TTS have been performed by \citet*{1984ApJ...277..725C}, \citet{1985ApJ...293..575C}, and \citet{1997ApJ...482..465G}, among others. These show line profiles that are broad ($\sim300$ \kms), with a central narrow absorption and a wide blueshifted absorption, superimposed to the broad emission. Observations of \mgii\ in cool stars \citep{1979ApJ...234.1023B} show that the central absorption can be produced either by ISM absorption or non-LTE effects in the stellar chromosphere. The blueshifted absorption indicates the presence of a cool outflow \citep{1997IAUS..182..417C}.

Most of the observational evidence for the Magnetospheric Accretion Model (MAM, the idea that the optical and ultraviolet characteristics of TTSs are the result of magnetically channeled accretion) comes from optical line and continuum emission. In the current picture (\citealt*{1994ApJ...426..669H}; \citealt*{1998ApJ...492..743M}; \citealt*{1998AJ....116..455M}, \citealt*{2001ApJ...550..944M}) it is assumed that the strongest optical lines (the members of the Balmer series, \nad, the \caii infra-red triplet, \hei) form in an accretion funnel. The blueshifted absorptions are the result of a cool outflow that, in general, does not produce emission. The model does a good job explaining the general characteristics of the data as a whole, although individual observations of individual stars still have unexplained features. \citet{1994AJ....108.1056E} used optical spectra of a sample of 15 CTTSs to confirm some of the magnetospheric predictions: blueward asymmetric emission lines and redshifted absorption components at typical free-fall velocities that are naturally explained by magnetospheric infall. However, a study of 30 TTSs by \citet{2000AJ....119.1881A} (AB2000) shows that \hal\ lines are symmetric, with broad wings often extending to $\pm500$ \kms. Traditionally, magnetospheric models have had a hard time accounting for this high-velocity emission, as the line widths depend only on the infall velocity of the gas, which for typical CTTS parameters is around 200 - 300 \kms. Furthermore, due to occultation effects, the models predict larger and more frequent asymmetries than those observed. A new set of models \citep{2001ApJ...550..944M}, incorporating stellar rotation (a small effect in general) and line broadening (dominated by the linear Stark effect in the case of \hal) can account for most of these discrepancies, as it predicts broad and fairly symmetric Hydrogen lines. The models naturally predict that lines such as \mgii\ should be much narrower than \hal\ (given that \mgii\ is not affected by the linear Stark effect), assuming that they are formed in similar regions.

Different regions of the funnel flow are expected to emit differently, and certain optical lines show evidence of multiple emission regions. \citet{1996ApJS..103..211B} show that \hei\ and \caii\ exhibit profiles typically with narrow (NC) and broad components (BC). AB2000 concludes that the Balmer lines and the BC  of other lines are mainly formed in the magnetospheric flow, with the NC formed close to the star.  The NC of the \caii\ and \hei\ lines are thought to be formed partly at the stellar surface, which is perturbed by the infall of material, and partly in the normal stellar chromosphere. 

\citet{1996ApJ...470..537M} argues that \mgii\ should be one of the most important coolants in the magnetospheric flow. This strong doublet provides us with another observational constraint that models should obey. If formed collisionally at low densities, the \mgii\ ion has a maximum population at 13000 K \citep{1992ApJ...398..394A}, although in the sun the self-reversed core is formed at $\sim7000$ K \citep*{1981ApJS...45..635V}. Being a resonant transition, \mgii\ is very sensitive to cool wind absorptions. In addition, the optical depth of \mgii\ k at line center (for a line with FWHM$\sim300$ \kms\ and assuming solar metalicity, with all the Mg as \mgii) is $6\ 10^{-19} \times N_H$, where $N_H$ is the hydrogen column density. The models by \citet{1994ApJ...426..669H} predict volume densities in the funnel flow larger than $10^{10} \rm{cm^{-3}}$ over distances of a few stellar radii. Therefore, if formed in the accretion funnel (as \hal\ is believed to be), the \mgii\ resonant lines, like the Balmer lines, should be very optically thick.

In this paper we present an analysis of GHRS data in a 40 \AA -wide spectral window centered around 2800 \AA, for a sample of eight CTTSs (BP Tauri, T Tauri NS, DF Tauri AB, RW Aurigae ABC, DG Tauri, DR Tauri, RY Tauri, and RU Lup) and one WTTS (HBC 388, also known as V1072 Tau and NTTS 042417+1744). These data are part of the same set of data in which Paper I is based. One of the defining characteristics of TTSs is their strong variability (photometric and spectral) on timescales of hours and days \citep{1993prpl.conf..543B}. Therefore, to make a comparison between the optical lines and the \mgii\ lines meaningful, simultaneous observations in the optical and UV must be arranged. For four of the stars we obtained nearly simultaneous (within 24 hours) optical data. The optical data span some of the strongest visible lines in the TTS spectra, including \hal, \hbeta, the \caii\ infrared triplet (IRT), \hei, and NaD. Such simultaneous multi-wavelength coverage is one of the best ways to understand the characteristics of the line formation region. Our purpose is to examine the data in the context of the MAM. As shown in the references, careful models of funnel flow emission and wind absorption have been published elsewhere and we will not repeat that work here.

The remainder of this paper is organized as follows: we will briefly describe how the optical and UV observations were taken, and provide an overview of the data. In Section \ref{anal} we will analyze the data, specially the \mgii\ resonance lines. We begin with a discussion of the narrow absorption feature present in these lines. In Section \ref{doublet} we study the kinematic characteristics of the \mgii\ emission components and compare them to the Balmer lines. In the next section we will discuss the relationship of the \mgii\ lines with the accretion phenomenon. Section \ref{wind} has a study of the wind characteristics in our stellar sample. At the end, our conclusions.

\section{Observations}
The characteristics of the observed targets are summarized in Paper I. All (except RU Lupi) belong to the Taurus star formation region, and we assume that all are at a mean distance of 140 pc from the Sun. They do not comprise a homogeneous sample in accretion rate: while RY Tau has low accretion rate ($\sim 10^{-9} \msun \ yr^{-1}$), and BP Tau, T Tau have average accretion rates ($\sim 10^{-8} \msun \ yr^{-1}$), the remaining stars have very large accretion rates ($\sim 10^{-7} \msun \ yr^{-1}$).

\subsection{UV Observations}

We obtained these data using the Goddard High Resolution Spectrograph (GHRS; see \citealt{1994PASP..106..890B}), one of the first generation instruments aboard the Hubble Space Telescope. The details of the reduction process have already been described in Paper I. Here we summarize them.

These data were obtained in GO programs executed in
cycles 3 and 5, as well in GTO observations in cycles 2 and 3.
Table~\ref{tab_log_mg} presents the log of observations. The observations before 1994 were made using the aberrated optics; later observations used the COSTAR corrective optics. The pre-COSTAR data are not seriously affected by the mirror aberrations, primarily because the lines are significantly broader than the instrumental width.

The observations were obtained through the large science aperture,
which projects to a 2~arcsec square on the sky pre-COSTAR, and to 1.74~arcsec post-COSTAR. On-target observations were summed using comb-addition (see Heap et al.\/ 1995). Each pixel corresponds to $\frac{1}{4}$~diode width; the instrumental resolution is about 4.4~pixels. The large science aperture subtends 8 diodes, or 32 pixels, in the dispersion direction. The instrumental resolution is constant with wavelength and is the same for each individual observation and the co-added frame. The resolving power is 30000 at 2800 \AA. The statistical uncertainty on the wavelength scale is 0.6$\pm$2~km~s$^{-1}$. There may be systematic offsets of up to $\pm$6~pixels ($\pm$12~km~s$^{-1}$) due to the target centering process, primarily for the pre-COSTAR observations. Further offsets may result from non-uniform filling of the aperture in the case of extended emission.
All readouts from a particular observation were interpolated to a common linear wavelength scale and summed. An effect of the interpolation is to smooth the data, and we scaled the error vectors accordingly.

\subsection{Optical Observations}

In order to make a direct comparison between the traditional optical
diagnostics and the UV lines, we arranged high resolution
optical spectra to be taken at the same time as the HST spectra.
Since T Tauri stars have been shown to be variable on the timescale of
a few hours in both the optical and UV, it was essential that they be
taken as closely together as possible.  

We arranged for the ground-based observations
once we knew when the HST observations were scheduled,
taking our chances that the stars would be observable from and the
appropriate instrument would be available at the Lick Observatory.
This strategy was successful for the four HST observations between
July 30 and August 10, because the Hamilton echelle was in use on the
3-m Shane reflector and Taurus rose slightly before morning twilight.
We were not able to arrange simultaneous observations for the 1995-1996 program. A log of the optical observations is shown in Table \ref{table_opt_log}. Some of our spectra were obtained with an interval of as little
as 8 hours between the UV and optical spectra, but more typically
12-15 hours.  We did not ask the observers to change the spectrograph
settings they were already in (which would also have entailed a new
set of calibration spectra).  We are deeply grateful to the observers who agreed to take these
observations.

The Hamilton is a cross-dispersed echelle with many orders per
exposure, so we knew that at least some of the interesting optical
lines would be in the format.  The observers sent us the raw spectra
and calibration exposures, and we used our standard echelle software
to produce our final spectra.  The absolute wavelength scale was
obtained as usual from ThAr spectra.  Reduction procedures have been
described by \citet{1994PhDT........16V}. The spectra were not flux-calibrated.

\section{General Characteristics}

\subsection{UV Data}
\begin{figure}
\plotone{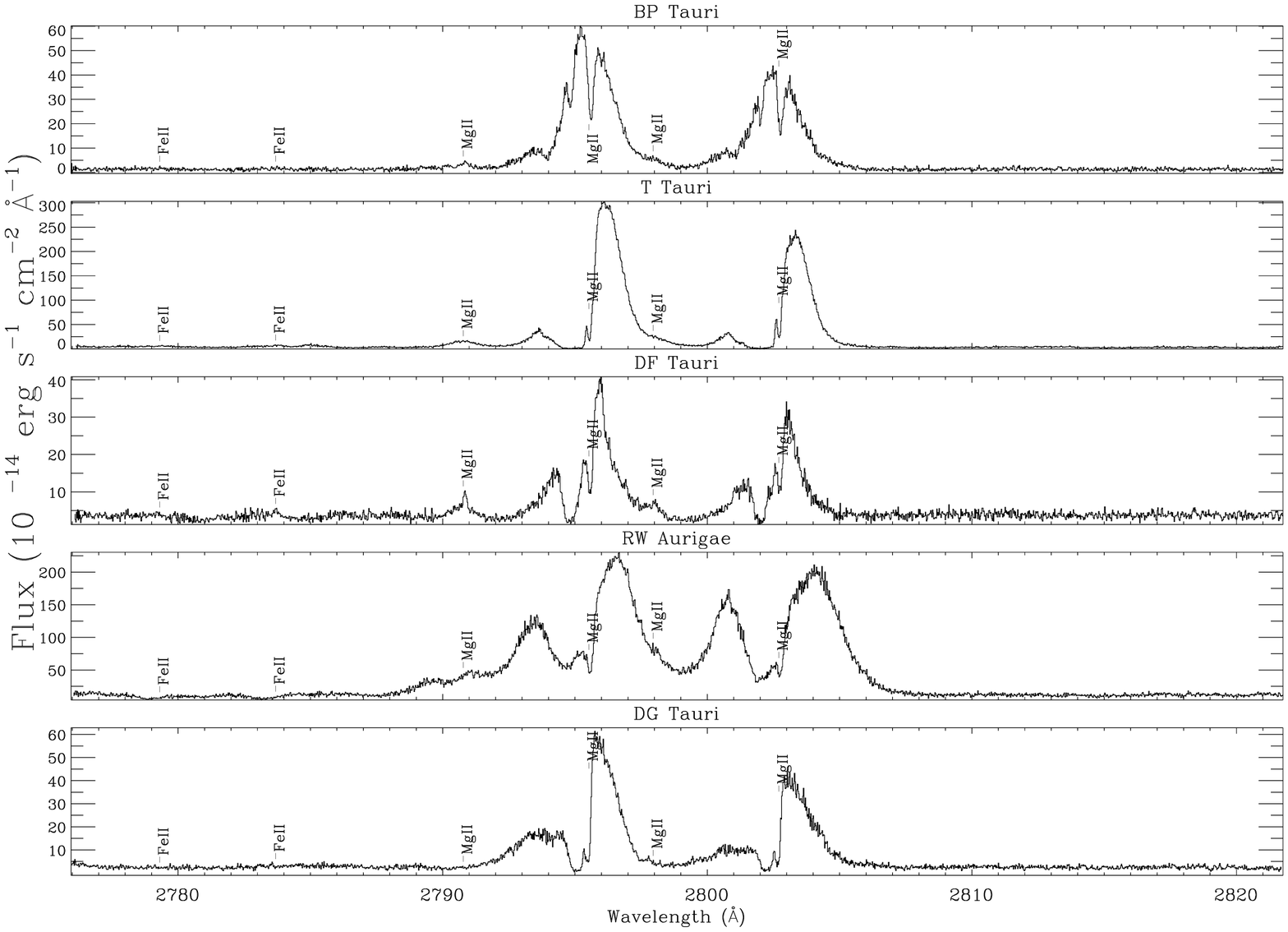}
\caption{Spectra centered around 2800 \AA. This region contains the \mgii\ doublet. For this plot, both observations of DG Tau have been averaged.\label{mgii_spec_1}}
\end{figure}

\begin{figure}
\plotone{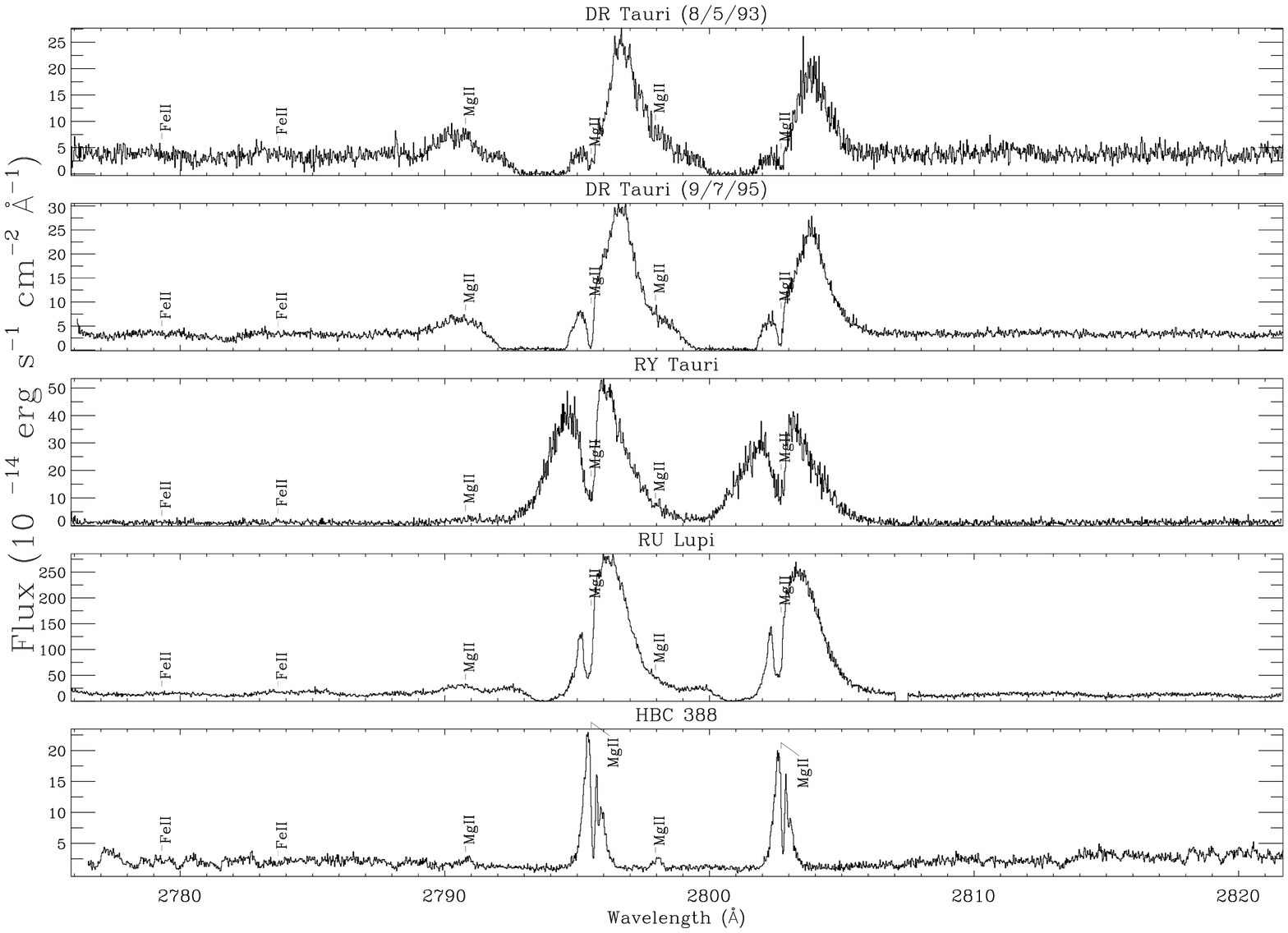}
\caption{Spectra centered around 2800 \AA, cont. The two observations of DR Tau in 1995 have been averaged.\label{mgii_spec_2}}
\end{figure}

Figures \ref{mgii_spec_1} and \ref{mgii_spec_2} show our ultraviolet data. All the spectra are in the stellar rest frame, with an air wavelength scale. For each spectrum we have marked FeII (2779.30 and 2783.69 \AA), the \mgii\ resonance doublets, and the subordinate \mgii\ lines belonging to the UV3 multiplet (2790.78 and 2797.96 \AA). The dead diodes for the RU Lup observations have been set to zero. For DG Tau and DR Tau (9/7/95) we have averaged observations taken on the same day. The spectra show wide \mgii\ emissions, with narrow and broad absorptions superimposed to them. A weak continuum is also present in most stars. In some stars, the members of the UV 3 multiplet are clearly seen blended with the \mgii\ resonance lines. We observe weak wind absorption in the subordinate \mgii\ line (2790.78 \AA) for RW Aur, DR Tau and perhaps DG Tau. This line is likely pumped by the strong resonance \mgii\ k line (\citealt*{2000A&A...357..951E}; \citealt*{2000AstL...26..225L}; \citealt*{2000AstL...26..589L}; \citealt*{LamzinonDF}).

\subsection{Optical Data}
\begin{figure}
\plotone{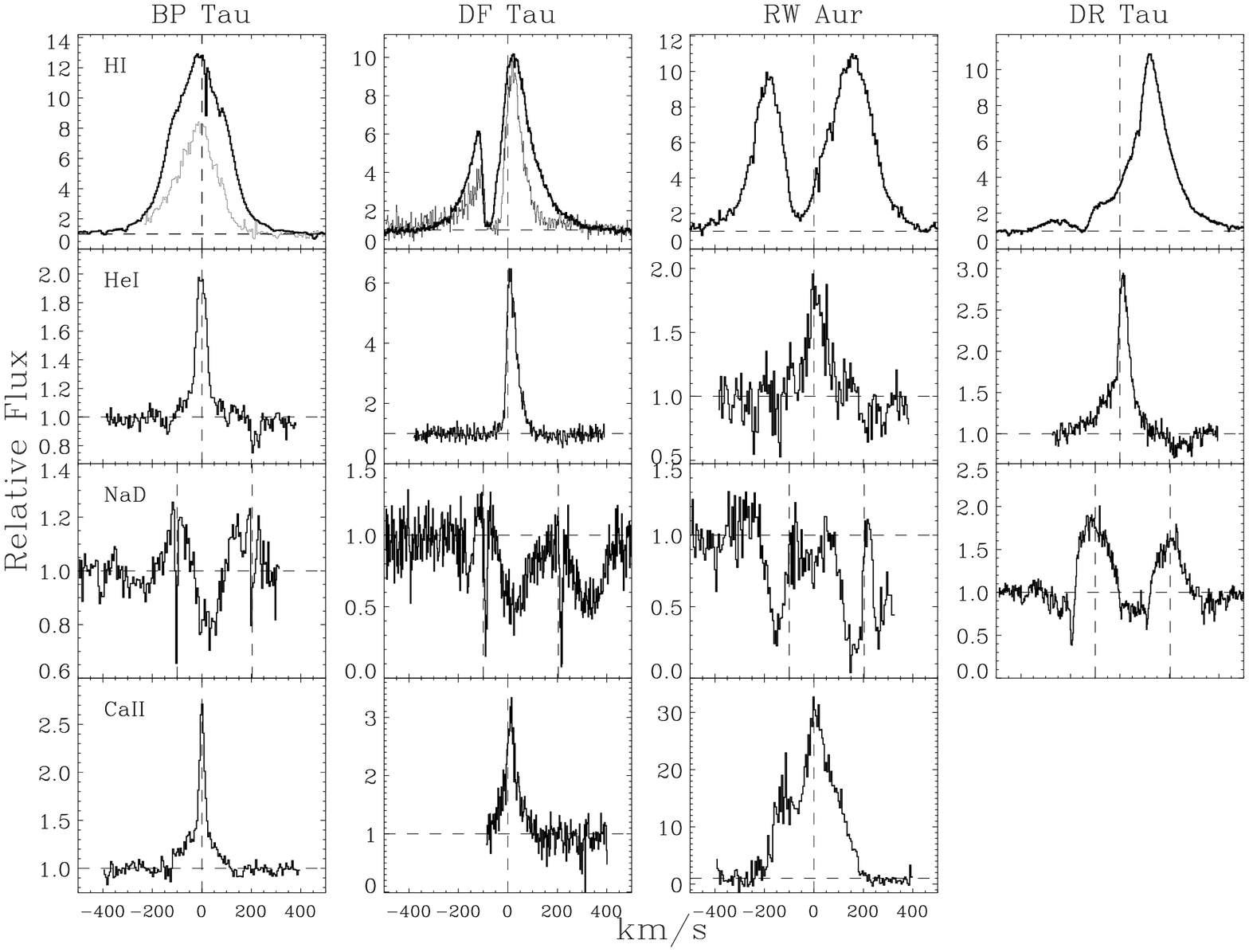}
\caption{\label{op} Strong optical lines. For H I, the thick line indicates \hal\, and the thin line indicates \hbeta. We do not have observations of \hbeta\ for RW Aur or DR Tau. We do not have Ca II for DR Tau. For the NaD lines, the spectra have been corrected for the sky lines, by fitting and subtracting a Gaussian to the emission. In the case of DR Tau and RW Aur, such procedure obliterates the weak ISM absorption. Only the bluest member of the Ca II triplet is shown. Zero velocity is indicated with a vertical dashed line for all lines.}
\end{figure}

Extensive analyses of the optical spectra of CTTS have been published elsewhere (e.g. AB2000), and therefore here we will concentrate on those lines that can be related to the UV diagnostics. In our optical spectra, the strongest features are \hal\ (6562.80 \AA), \hbeta\ (4861.32 \AA), \hei\ (5875.63 \AA), NaD (5889.95, 5895.92 \AA) and the \caii\ infrared triplet (8498.02, 8542.09, 8662.14 \AA). We have spectra of \hal, \hei, and NaD for the four stars. We have spectra of \hbeta\ for BP Tau and DF Tau, and spectra of the bluest member of the IRT for BP Tau and RW Aur. These emission lines are shown in Figure \ref{op}, in which the accretion rate increases from left to right. For DF Tau we also have all the three members of the IRT (only the bluemost is shown). The lines for DR Tau are the average of two observations separated by 24 hours. We do not have the same lines for all stars due to the fact that they were obtained by other observers, and, in order to minimize the impact in their observations, we did not ask them to change their settings. As we are interested mainly in comparing these lines with the UV lines, here we present a simple overview of the data. 

The \hal\ lines are broad (FWHM$> 200$ \kms). Three of them show two distinct peaks, separated by a blueshifted absorption. \hbeta\ is narrower (FWHM$< 170$ \kms) and weaker. None of the Balmer lines show redshifted absorption.
The \caii\ IRT  shows narrow and broad components. In the case of RW Aur, the components are so separated that two peaks are visible. The Paschen line Pa15 that is blended with \caii\ in some of the spectra analyzed by AB2000 is not observed in any of our profiles. The NaD line shows the characteristic filled absorptions.
The \hei\ line also shows narrow and broad components. In addition, BP Tau, DR Tau, and possibly RW Aur, show a redshifted absorption component, as has been documented previously (AB2000). In summary, the optical spectra simultaneous with the UV spectra are not unlike other spectra that have been observed and analyzed previously. This is an important point, as we will generalize from this observations to the usual behavior of the star.

Using the data from Paper I, we can compare optical lines with simultaneous UV lines. \citet{1996ApJS..103..211B} argue that the part of the narrow components due to accretion in the \hei\ and \caii\ lines, is formed in the chromosphere of the star, next to but outside the footpoints of the accretion column. The \civ\ and \siiv\ resonance doublets are also believed to be formed close to the star, where the temperatures are highest. In Paper I we found that the morphology of the \civ\ lines is very different from that of the \hei\ or \caii\ lines: as Table \ref{all_lines} show, the latter lines tend to be narrower (FWHM less than $\sim$200 \kms, compared to FWHM$\sim$200 \kms\ in most cases for \civ) in those stars for which we have simultaneous optical and UV spectra. For the two stars in which we perform decompositions of the \civ\ lines (BP Tau and DF Tau) we do not find any conclusive relationship between the width, velocity centroids or flux of the Gaussian components of these lines and those of \hei\ and \caii. This is not surprising given the very different thermal regimes both sets of lines sample.

\section{Analysis \label{anal}}
We have measured the parameters of \mgii\ resonance lines and of the optical lines indicated in Figure \ref{op}. The results are shown in Table \ref{all_lines}. For each measurement we have used the noise vector provided with the data to obtain an error estimate. In general, we have measured fluxes, full widths at half-maxima, and velocity shifts, by fitting a Gaussian or a combination of Gaussians to the profiles, using a Marquardt routine. The \mgii\ lines are modeled as the sum of one emission component and one or more absorption components. We choose to analyze the profiles in terms of Gaussian components because they show symmetric wings. The Gaussian decomposition does a good job in describing the shape of the red wing and the far blue wing of the profiles. In Table \ref{all_lines} we give the parameters for the emission components only, after subtraction of the members of the UV 3 multiplet, if possible. The members of the multiplet are clearly separated by the fitting routine in most cases. For DG Tau, RY Tau, and RU Lup we find that one Gaussian emission component correctly reproduces the emission, even at the wavelengths affected by the UV3 multiplet. For these stars no attempt to subtract the multiplet has been made. 

For the wide blueshifted absorption we indicate its velocity and the FWHM, if the absorption can be fit by a single Gaussian (which is only possible in some cases). For all the blueshifted absorptions we indicate the ``terminal velocity'', the velocity at which the absorption reaches the continuum (either the real continuum or the \mgii\ line wing). We model the Balmer lines as a strong emission with superimposed absorptions. The \hei\ and \caii\ lines are modeled as the sum of a BC and a NC. For each optical profile, we measure the center and the full width at half maximum (FWHM) of the emission components. This procedure is not applicable to the \nad\ lines.

\subsection{The \mgii\ Resonance Doublet}

\subsubsection{Narrow Central Absorption}

As Figures \ref{mgii_spec_1} and \ref{mgii_spec_2} show, there are narrow central absorptions superimposed to the wide resonance lines of \mgii. Observations of the \mgii\ resonance lines in cool stars \citep{1979ApJ...234.1023B} show strong central core reversals and ISM features. When it is assumed that the absorptions are due only to ISM absorption (by which we mean any absorption by material not related to atmospheric processes in the vicinity of the star: i.e., material in a circumstellar or intervening cloud) it is not uncommon to find absorption components whose characteristics do not correspond to those predicted by models of the Local Interstellar Cloud (LIC) or the G cloud \citep{2000ApJ...542..411W, 2001AJ....121.2173B}. These extra components are perhaps due to other smaller and yet to be described clouds. On the other hand, given that the characteristics of the line formation region in TTSs are not very well known, it is not clear what one would expect in terms of self-reversals in the core of the \mgii\ line. It is for this reason that we have decided to examine the question of the nature of the narrow central absorption present in all our spectra, which has in the past been identified with ISM absorption \citep{1997ApJ...482..465G}. Both possibilities (ISM absorption vs. intrinsic NLTE self-reversals) are not mutually exclusive: for example, the core of a NLTE self-reversal may be affected by ISM absorption. We conclude below that the central absorption is impossible to understand as being due just to absorption of interstellar material in the line of sight.

Table \ref{table_ism} lists measurements of the velocity of the line (in the stellar rest frame and with respect to the \mgii\ lines) and its FWHM. We have obtained these parameters by simple Gaussian fittings to the profiles, including the wind absorption. The latter is very saturated in most stars and we have not attempted a perfect fit to the profile, only to the region with the narrow absorption. The table also includes the radial velocity of the star, the predicted velocity of the ISM\footnote{http://casa.colorado.edu/$\sim$sredfiel/ColoradoLIC.html.}, either the LIC or the G cloud, as modeled by \citet{2000ApJ...534..825R}, and the difference between the two.

Measurements of the equivalent width (EQW) of the observed narrow absorptions (also shown in Table \ref{table_ism}) yield a wide range of values, from 0.1 \AA\ for T Tau, to 0.3 \AA\ for RU Lup and 0.6 \AA\ for RY Tau. These are uncertain measurements, given that the `continuum' is, most of the time, the rapidly varying red wing of the blueshifted absorption. After division by the Gaussian fits to the emission component (Section \ref{doublet}), the narrow absorptions in the k and h lines reach similar depths (which means that the lines are saturated). No correlation of EQW with inclination or accretion rate is observed. Notice the variability in the EQW of DR Tau between the two epochs, which is larger than the variability in the emission line flux.  Additionally, note that RY Tau is the only CTTS in our sample for which no wide, blueshifted absorption is observed (Section \ref{wind}). The central absorption in this star (the widest absorption in the sample) is asymmetric, rising more gently on the blue than on the red side (which explains the large negative velocity in Table \ref{table_ism}). These observations point to the fact that we are seeing the effects of a low velocity wind in the narrow central absorption feature of RY Tau. \citet{2001ApJ...551..413R} have observed ISM absorption, using STIS/HST in the \mgii\ lines of 18 stars from the Hyades, which is in the general direction of the Taurus cloud. In their work, they characterize a new cloud, the Hyades cloud, between the LIC and the Hyades. From their paper, we measure equivalent widths of the order of 0.1 \AA. This is the lower limit of the equivalent widths measured in this work.

From Table \ref{table_ism} the velocities of the narrow absorption coincide with those predicted by the LIC model in the case of BP Tau, RW Aur, and DR Tau. Also, \nad\ profiles of BP Tau and DF Tau show that for these stars, the velocity of the absorption in \mgii\ is within 5 \kms\ of the absorption in \nad.  We do not see any correlation between the velocity of the absorptions and velocity centroids in other lines of the same star, either for the \mgii\ emission components or the \htwo\ lines that were measured in Paper I. We should point out that due to targeting errors in our data, there may be systematic offsets in velocity of up to 12 \kms\ between different stars. These systematic errors will affect the measured velocity of the narrow absorption. Given the possibility of additional clouds and the systematic errors, the match (or lack of it) in velocity space between the measured and the predicted velocities is not very significant.

A different handle on the origin of the narrow absorption is provided by the line widths.  \citet{2001ApJ...551..413R} measure Doppler parameters for the ISM absorption that vary between 2 and 4 \kms. Separation in velocity between the Hyades cloud and the LIC can be as large as 10 \kms. Fitting a Gaussian to their observed ISM lines, it is possible to obtain k line full-width at half maxima of about 20 \kms\ (including the correction for the instrumental resolution of the GHRS). These widths are again a lower limit to our observed widths.

On the other hand, the width of the absorption in HBC 388 is similar to that of nearby cool stars like $\beta$ Cet, $\lambda$ And, and $\mu$ Vel \citep{1979ApJ...234.1023B}. In these cool stars, the absorption is thought to be due to NLTE self-reversals. The HBC 388 absorption is the narrowest in the sample which suggests that the widths of the absorptions are determined by a process related to the accretion of the CTTSs. If we are seeing a self-reversal due to material in the funnel flow (assuming that this is the origin of the \mgii\ emission), the width will be determined by the geometry of the accretion sheet or tube. In this case one would indeed expect a wider absorption in CTTSs than in WTTSs, given the large range of line of sight velocities available. On the other hand, no correlation is observed between the width of the lines and the accretion rate or the inclination.

A comparison with measured $v \sin i$ (taken from \citealt{2000ApJ...539..815J}), does not help to decide the issue. Since the absorption lines are saturated, the effect of the rotation is likely to be small. Projected rotational velocities for our stars range from 10 to 20 \kms. In this case the Gaussian kernel of the rotationally broadened profile has  a FWHM of only 5 to 10 \kms, very small compared to the line width. We do not observe any correlation between measured $v \sin i$ and the FWHM of the absorptions.

In summary, the narrow central absorptions of the stars in our sample have larger EQW and larger FWHM than the ISM absorptions (due to the LIC and the Hyades cloud) observed in stars on the same line of sight. It is not impossible that other, farther clouds may exist that contribute to the absorption. Furthermore, the Taurus region itself may produce absorption in \mgii.  To produce a 20 \kms\ FWHM absorption, the cloud material should have a Doppler parameter $b\sim10$ \kms, much larger than other values observed in the ISM. In addition, the fact the HBC 388 has the narrowest absorption in the sample lead us to conclude that most of the width of the observed narrow absorption should be due to non-LTE processes in the T Tauri system itself. The lack of correlation of the line parameters (FWHM and EQW) with inclination or accretion rate indicates that the accretion process is not narrowly confined in angle and that the decay of the \mgii\ source function is insensitive to changes in accretion rate by over an order of magnitude.

\subsubsection{Emission Profile: Kinematics and Comparison with Balmer Lines  \label{doublet}}

As the flux ratios in Table \ref{all_lines} show, none of the emission components of the \mgii\ lines are optically thin. For some stars, the k line is significantly narrower than the h line (RW Aur, DG Tau, RU Lup), and, in all the stars except RW Aur, the k line flux is brighter than the h line flux. The fact that the \mgii\ k line is generally stronger than the h line may be due a geometrical effect: the more optically thick k line will be optically thick over a larger area. Of course, this assumes that velocity and source function gradients do not negate such a simple argument.

In general, there is a difference between the centroids and widths of the emission for each member of the \mgii\ doublet, and we see the stars divide in two groups: on the one hand DG Tau, RU Lup, and RW Aur show a k line narrower and blueshifted with respect to h. The rest of the stars show small differences (up to 10 \kms) between both components, both in the width and the velocity of the centroids. With such a small sample is not clear if this behavior is simply the result of differences in the source function inherent in each source, or it reflects different formation places in the stars. This is definitely one of the behaviors that any model of the region should explain, akin to an explanation of the Balmer decrement in the optical range.

\begin{figure}
\plotone{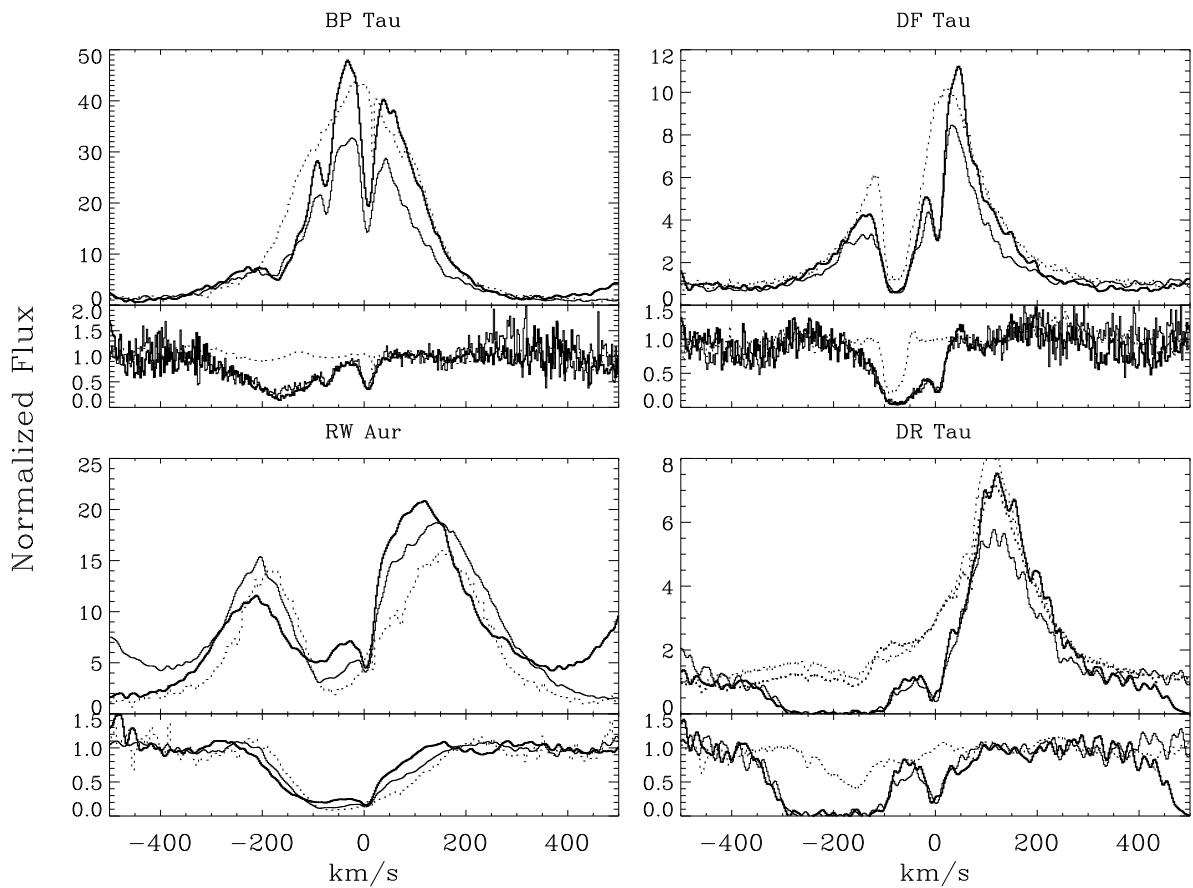}
\caption{\label{mgii_1}The \mgii\ doublet and \hal. The solid thick (thin) line shows the \mgii\ k (h) line. The members of the UV 3 multiplet have been subtracted. The dotted line is \hal. All the spectra have been continuum normalized. The vertical scale for the \hal\ line has been stretched to match the red wing of the \mgii\ k line in all the stars, except in RW Aur (for this star the blue wing of the h line is visible in the thick profile). The bottom panels for each star show the result of dividing by a single Gaussian representing the emission component. In the case of RW Aur, the red end of the k line merges with the blue end of the h line. In the case of DR Tau, the two dotted lines show two \hal\ observations separated by 24 hours. To obtain the bottom panel for DR Tau, we divide by the emission components of the average \hal. For this star the k line is affected at large positive velocities by the very broad absorption in the h line}
\end{figure}
\begin{figure}
\plotone{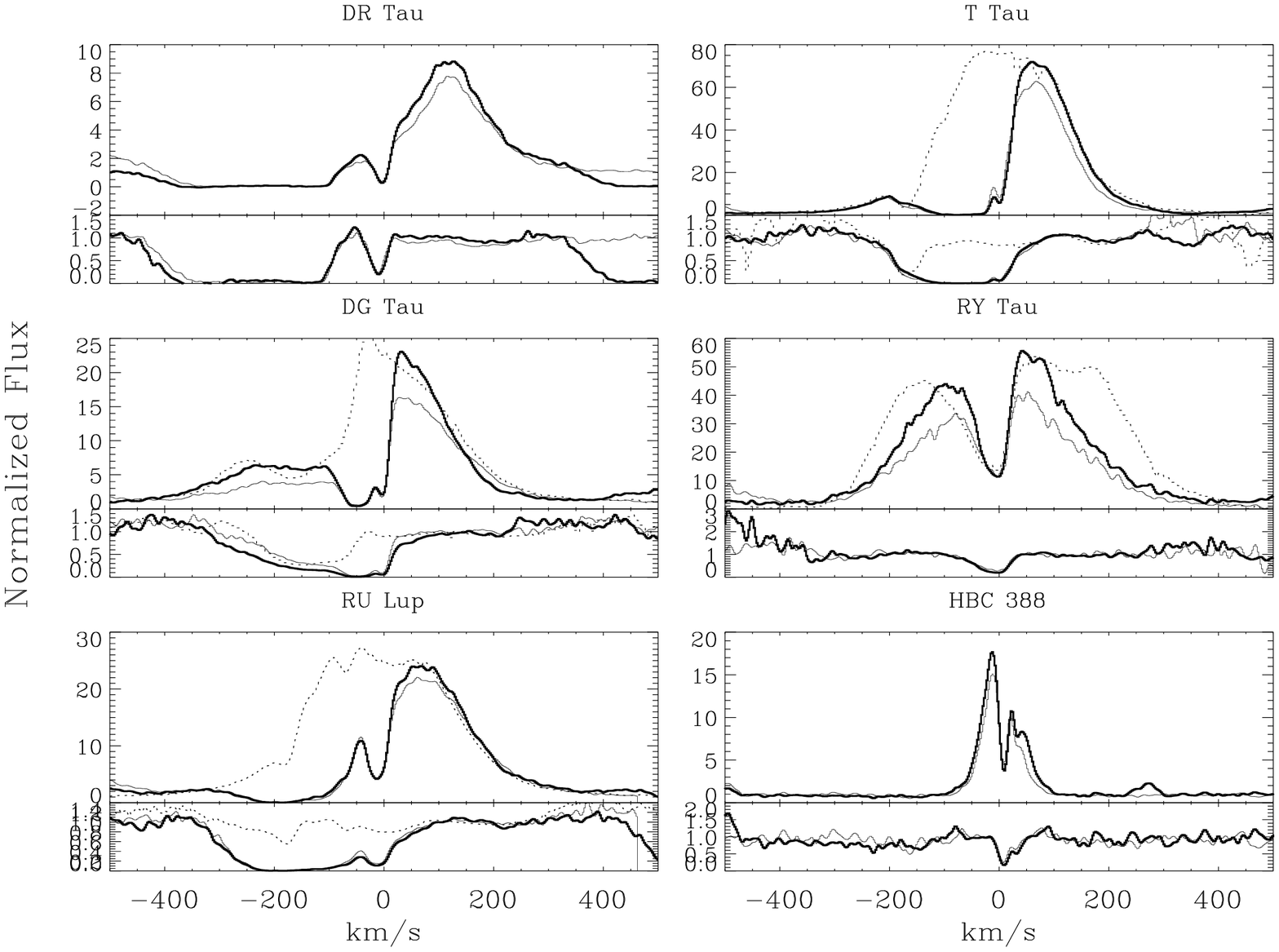}
\caption{\label{mgii_23}The \mgii\ doublet for stars without simultaneous optical data. All the spectra have been normalized. The bottom panels for each star show the result of dividing by the emission components. For RY Tau, a Gaussian fit to the emission component of \hal\ is not possible as the wings are not symmetric. The \hal\ line in HBC 388 is very weak, and we do not show it. For this star, we have not subtracted the red members of the UV3 multiplet.}
\end{figure}

Figure \ref{mgii_1} shows the \mgii\ doublet and the \hal\ line on a common velocity scale for those stars for which we have simultaneous optical and ultraviolet data. Figure \ref{mgii_23} shows the \mgii\ profiles for stars with no simultaneous optical observations. The second epoch of observations for DR Tau does not have an \hal\ profile. \hal\ lines for T Tau, DG Tau, and RY Tau have been published by some of us previously (\citealp*{1995AJ....109.2800J}, AB2000) and represent average profiles. The \hal\ line of RU Lup is an individual observation graciously provided by Celso Batalha. The \mgii\ profiles have been normalized to their local continua. To make the comparison between the two sets of lines meaningful, we stretch the vertical scale of \hal\  to match the red wing of the k line if possible. For RW Aur, such matching is not possible and so we scale the \hal\ line to the blue peak of the \mgii\ h line. The contributions of the UV3 multiplet to the \mgii\ k line have been subtracted when possible.

The similarity between the \mgii\ lines and the \hal\ lines is remarkable. In most stars, the wings of the lines are well matched, as are the absorption features. The depth of the blueshifted absorption tends to be deeper and wider in \mgii\ (a resonance line) than in \hal, as discussed in the introduction. The similarity of the emission components in \hal\ and \mgii\ strongly suggests that both have a common origin and are affected by the same processes. As Table \ref{all_lines} shows, the line width is similar (but not identical) in \mgii\ and \hal\ for all the four stars for which we have simultaneous optical and UV spectroscopy. The most glaring exception is RY Tau, which we discuss below.

The comparison also highlights the limitations of modeling the emission component of the \mgii\ line as a single Gaussian. The emission component of the \hal\ line in RY Tau or RU Lup cannot be fitted by a single Gaussian: for the former the profile is not symmetric, for the later the core of the line ($|$v$|<$100 \kms) is not Gaussian. The procedure outlined here works in part because the cores of the \mgii\ lines are, in general, not present, and the wings are symmetric. For our purpose (to compare \hal\ and \mgii, and to analyze the kinematics of the wind), the single Gaussian approach is justified.

 \begin{figure}
\plotone{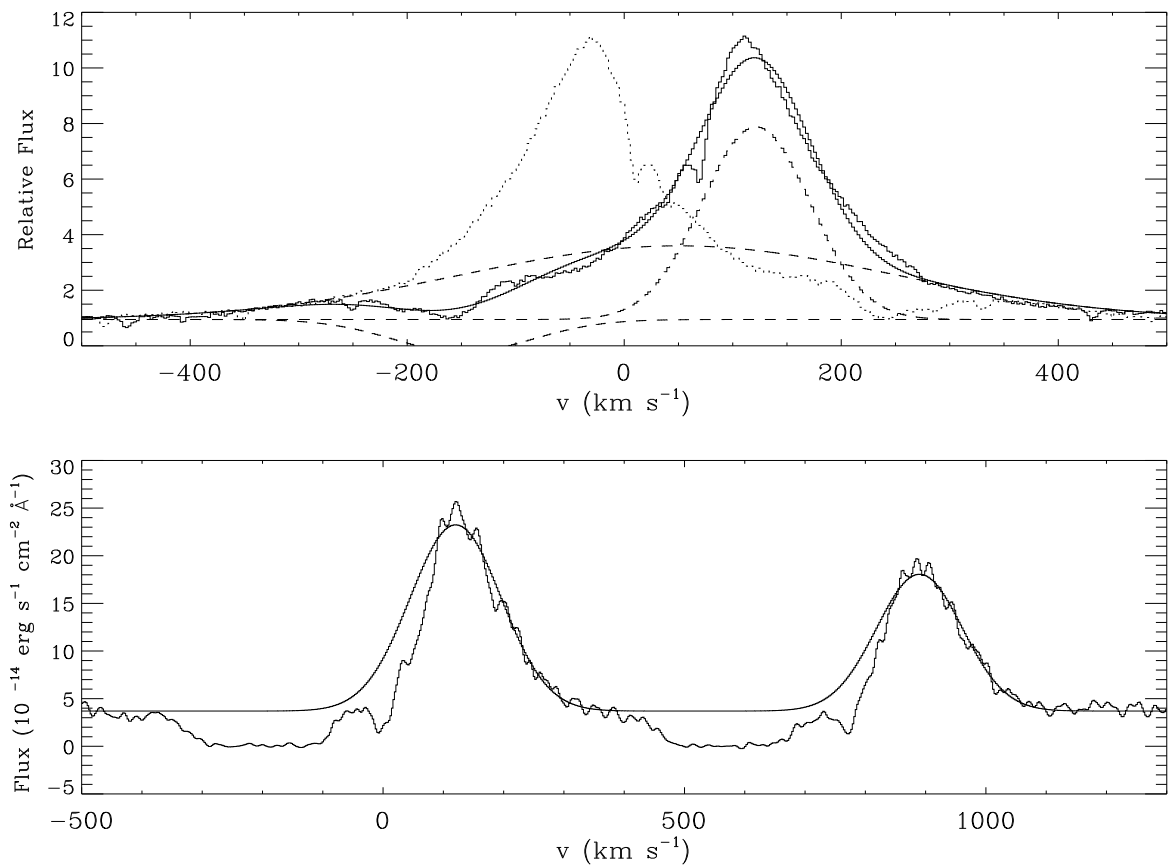}
\caption{\hal\ and \mgii\ profiles for DR Tau. The solid irregular line is the observed profile.  The solid smooth line is the Gaussian fit. Top: \hal. The dotted line is the average of the observed profiles, reflected around 80 \kms. The dashed lines are the fit components. Bottom: \mgii, first epoch. We have subtracted the components from the UV3 multiplet. The velocity scale is centered on the k line. The Gaussians are fit to the reflected \mgii\ profiles. These are not shown here, but they are traced by the total fit.\label{dr_fit}}
\end{figure}

For DR Tau, we have \hal\ observations contemporary with the first epoch of UV observations (8/5/93). As Figure \ref{mgii_1} shows, the blue wings of the \hal\ line change (perhaps due to variable wind absorption), while the red wings remain almost constant. A detailed analysis of the temporal behavior of DR Tau appears elsewhere \citep*{AlencaronDR}. The red wings of \hal\ follow the \mgii\ k line very closely. We have used the fact that the far wings of the average \hal\ line are symmetric to fit a low, very broad component (VBC) to them (see Figure \ref{dr_fit}, top plot).  We believe that the \hal\ line can then be described in terms of three Gaussian components: the low, very broad component (FWHM=$470$ \kms), a broad (BC) redshifted emission component (FWHM=$117$ \kms) and a blueshifted absorption (FWHM=$150$ \kms). For the \mgii\ profiles, a number of different decompositions are possible. It seems unlikely, however, that a very broad component is present. From Figure \ref{mgii_1}, the red wing of the h line drops very quickly to its continuum level (the red wing of the k line is affected by the wind absorption of h). The same is true for the second epoch of DR Tau observations. To fit the lines, we reflect the profiles around their center of symmetry and to this composite profile we fit a Gaussian. As the figure shows, such a procedure reproduces the red wing of the line very well. There is no real evidence for a VBC in \mgii.

HBC 388 is a WTTS and, as mentioned in the introduction, its emission is unrelated to accretion processes. The FWHM of the emission components is $\sim70$ \kms. This star shows a redshifted absorption in both lines, at $\sim30$ \kms. We believe this is due to downdrafting material condensing out of prominences. Similar redshifted absorptions have been observed in \hal\ for solar flares \citep{1997ApJS..112..221J} and in WTTS \citep{1996AJ....111.2066W}. Assuming the specific intensity of \mgii\ is the same for HBC 388 and the CTTSs, the flux in the emission lines due to non-accretion processes should be in the ratio of the square of the radii. From the stellar radii quoted in Paper I, Table 1, we conclude that the accretion process dominates emission in \mgii. Here we should add the same caveat as in Paper I: HBC 388 is an early K and it is unclear how representative its atmosphere is for all the CTTSs in our sample. 

The origin of the \mgii\ lines in CTTS is not fully understood, but the similarity with \hal\ indicates that they come from the same region. Models to produce the line in an spherical \citep{1990ApJ...349..168H} or conical \citep*{1992ApJ...386..229C} wind tend to produce deeper than observed blueshifted absorptions in the \hal\ profiles. In addition, \citet{1995AJ....109.2800J} have shown that, for a sample of seven stars, the blueshifted absorption is uncorrelated with the rest of the line. As mentioned in the introduction, current models \citep{2001ApJ...550..944M} assume that, in general, all the emission component of the \hal\ line is formed in the funnel flow and the blueshifted absorption is the result of the T Tauri wind, which is assumed not to produce any emission. In these models, Stark broadening is necessary to reproduce the large widths and the symmetric appearance (AB2000 find symmetric wings in \hal\ for 80\% of their stars) of the \hal\ lines, specially in high accretion rate stars like DG Tau, DR Tau, and RU Lup. However, the resonance lines of \mgii are not affected by the linear Stark effect and is only weakly affected by the quadratic Stark effect, at least compared to \hal. Only in the case of DR Tau it seems that the very broad component observed in \hal\ is not present in \mgii, as would be expected from the Stark effect.  From the fact that their models cannot reproduce its shape, \citet{2001ApJ...550..944M} argue that the \hal\ emission from DR Tau comes mostly from the outflow. However, using a simplified model, \citet{AlencaronDR} show that by changing the temperature profile of the funnel flow and the inclination of the star, one can reproduce the narrow emission component. They argue that the breadth of the \hal\ BC emission may be due to magnetic turbulence, and it is redshifted due to the viewing geometry of the system. This could also be the case for the \mgii\ emission components, which are also redshifted.

The symmetry and width of the Balmer lines, and the similarity of the \hal\ profiles to \mgii\ suggests an atom independent broadening mechanism. Such mechanism may be supersonic Alfv\'enic turbulence, as has been suggested in the past (\citealp{1990MmSAI..61..707B,1997ApJ...474..433J}). This turbulence likely results from inhomogeneities in the accretion flow into the star. To complicate matters in this issue, \citet*{2001ApJ...551.1037B} have suggested that the wings of the \hal\ lines for $\sim50$\% of CTTS show indications of being formed in a hot coronal wind, which is also responsible for the broad component present in \hei. However, as \mgii\ is a resonance line, it is likely that this wind will emit over larger volumes and produce larger wings in this line than in \hal\ (See Section \ref{wind}). So, the picture that the observations presented here provide is that of a \mgii\ line formed in the funnel flow, with supersonic turbulence being the dominant mechanism in the formation of the wings for this line and for \hal.

The previous paragraph does not apply to RY Tau, an edge-on system with the lowest accretion rate in our sample ($2.5 \ 10^{-9} \msun \ yr^{-1}$, Paper I). For this star the \hal\ line is broader than \mgii\ and asymmetric. RY Tau is a type III variable (in the classification originated by \citealt{1994AJ....108.1906H}), which shows occasional obscurations by circumstellar dust. \citet{1999A&A...341..553P} and \citet{1995AJ....110.2369E} argue that at least part of the emission lines (\hal\ and \mgii, in particular) originate outside the region of obscuration, although it is difficult to see how emission at many AU can be responsible for the rapid variations (in timescales of days and less) observed in \hal\ for this star \citep{1995AJ....109.2800J}. In addition, \hal\ presents absorption components that appear and disappear in these timescales, preferentially at velocities of 0 and -100 \kms. In other words, it is not clear that non-simultaneous observations of these lines are comparable at all. High spatial resolution observations in the emission lines may help resolve the situation.

\subsubsection{Relation between Accretion Rates and Line Properties}

In the context of the MAM, the large brightness and large FWHM of the Balmer lines are believed to be due to the fact that that they originate mostly in the accretion funnel. It is therefore useful to see to what extent these lines are related to the measured accretion rates. A strong correlation has been observed between accretion rate and Br$\gamma$ \citep{2001ApJ...550..944M} which implies a correlation between the accretion spot size and the accretion rate. 

The \hal\ fluxes (from \citealp*{1974A&AS...15...47K}) and the \mgii\ fluxes are have similar order of magnitude, although for this sample, the \mgii\ fluxes show a larger spread for a given accretion rate, than the \hal\ fluxes. No correlation is observed between the line flux and the accretion rate. It is not clear that a correlation is expected. For a sample of 30 stars, AB2000 did show that a correlation exists between the veiling corrected \hal\ EQW and the accretion rate. However, the sample used in that correlation comprises mainly low- and medium-accretion rate stars and has only two stars in common with our sample: BP Tau and DF Tau. \citet{1998ApJ...492..743M} suggest that weak correlations between \hal\ and accretion rates are expected for accretion rates larger than $\sim10^{-8}\msun \ yr^{-1}$, as the rate of growth of the line flux decreases for these accretion rates. 

Even for the whole range of accretion rates, weak correlations should be the norm, given that the relationship between flux and accretion rate depends on the magnetospheric parameters of each star. Furthermore, accretion rates are uncertain, and known to change with time \citep{2000ApJ...539..834A}. We do not find any significant correlation between \mgii\ line flux, width, or centroid, and the stellar inclination. We do not find any correlation between wind EQW (as derived in the next section) and the \mgii\ emission flux. Any conclusions must be tempered by the fact that we are using a sample of only nine observations, which is biased to strong accreters.

\subsubsection{\mgii\ and Outflow Diagnostics\label{wind}}

As mentioned in the introduction, \mgii\ is expected to be very sensitive to the CTTS outflow. Such expectation is confirmed in Figures \ref{mgii_1} and \ref{mgii_23}, which show very strong absorptions in \mgii, even when such absorptions are weak or absent from the \hal\ profiles. For all stars the blueshifted absorptions are very similar in the h and k lines. The similarity of the normalized wind profiles in the h \& k lines for all the stars implies that the wind is optically thick. The bottom of the absorption then corresponds to the wind source function, which is zero or close to zero in all stars.

All the stars show blueshifted absorption. The RY Tau narrow central absorption has the largest EQW of the sample and is blue asymmetric. This suggests that a low projected velocity outflow is blended with the narrow central absorption. The star is almost edge-on and has the smallest accretion rate in our sample. Another almost edge-on object is DF Tau, with 70 times the accretion rate. For this star the wind is clearly visible. Below we explore the connection between the accretion rate and the EQW of the wind signature.

Multiple blueshifted absorptions are observed in BP Tau and DG Tau. In the latter, such absorptions are also present in \hal\ (with velocities -103 and -162 \kms) and perhaps \hbeta\ (AB2000). These stand in contrast to the (non-simultaneous) velocities measured in the blueshifted absorptions of the \mgii\ lines: -38 and -112 \kms. The difference is significant, in spite of the systematic velocity shifts in our data. The multiple wind signatures indicate that the wind emission or the source of the wind heating is episodic. It may also indicate the presence of more than one wind source, as has been argued by \citet{2001ApJ...551.1037B}.

In the stars for which we have simultaneous \hal\ observations, we observe either no wind in \hal\ (BP Tau), a narrow \hal\ absorption (DF Tau, DR Tau), or a wide \hal\ absorption (RW Aur). One could argue that narrow absorption in \hal\ (narrower than that in \mgii) is expected: the limits in the wind signature of \hal\ in velocity space are determined by the agent populating the line, either collisions or recombinations from the continuum. As the \hbeta\ profile for DF Tau shows (Figure \ref{op}) the high velocity limit is the same for \hal\ and \hbeta, which is consistent with a decrease in n=2 hydrogen population. Assuming that the wind accelerates outwards, the low velocity limit is likely due to thermalization close to the CTTS (in which the line populations are mainly determined by collisions), which occurs earlier for \hal\ than for \hbeta. \mgii\ h \& k, being resonance lines, are sensitive to larger wind volumes and/or smaller wind column densities. 

This simple picture base on atomic parameters cannot be easily generalized. In RW Aur, the \hal\ and the \mgii\ wind follow each other closely. In the low velocity end this may be taken as an indication that the wind is launched from a relatively low density region, given that \hal\ is not thermalized. The stars with no simultaneous optical data tell a more complicated story: the wind reaches the same terminal velocity in \mgii\ as in \hal\ for T Tau, smaller in RY Tau, and larger RU Lup. It may be that the relationship between the \hal\ wind and the \mgii\ wind reflects particular conditions of each star and not necessarily general characteristics. What is clear is that the \mgii\ resonance doublet presents a clear  outflow signature even when that signature is weak or absent from \hal. We should also remember that the wind strengths and terminal velocities are all variable and so non-simultaneous observations are not necessarily comparable.

\begin{figure}
\plotone{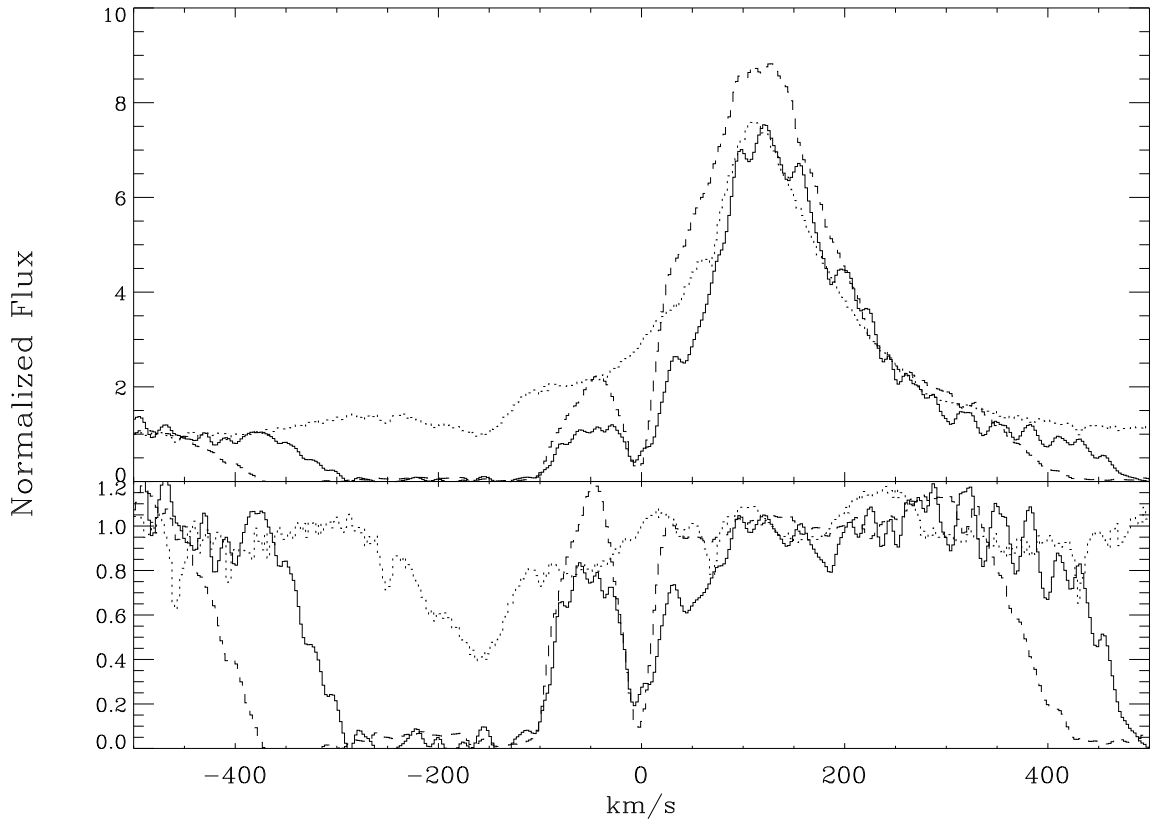}
\caption{\label{drtau_wind}Time variability in 
DR Tau. The solid (dashed) line shows the \mgii\ k line for the first (second) epoch of observations. The dotted line is the average of the two \hal\ observations. Notice how the terminal velocity of the wind (as measured by \mgii) increases from the first to the second epoch.}
\end{figure}

The two different epochs of DR Tau allow us to study the variability of the wind emission (Figure \ref{drtau_wind}). Between the two epochs there is change in the continuum (an increase of 15\% in the overall level), and in the \mgii\ flux (an increase of 20\% in the peak level). In the second epoch, the terminal velocity of the \mgii\ wind increases from $\sim350$ \kms\ to $\sim 450$ \kms. It is unclear whether this change in terminal velocity is due to increased density in the outflow or to increased terminal speed in the wind.

We do not observe any correlation between terminal velocity and accretion rate or inclination. However, a better indicator of the overall properties of the wind is the EQW, which simultaneously measures the width and depth of the line. Based on the similarity of \mgii\ with the \hal\ line, the failure of models to explain the Balmer lines assuming that they are created in an spherical wind \citep{1990ApJ...349..168H}, and the general success of the magnetospheric models in reproducing line fluxes, we assume that the emission component of the line is created in a different region than the wind. Making this assumption, the wind EQW is independent of the rest of the line. In particular, it does not need to be corrected for veiling, as it is usually done with measurements of photospheric lines of CTTS. 

\begin{figure}
\plotone{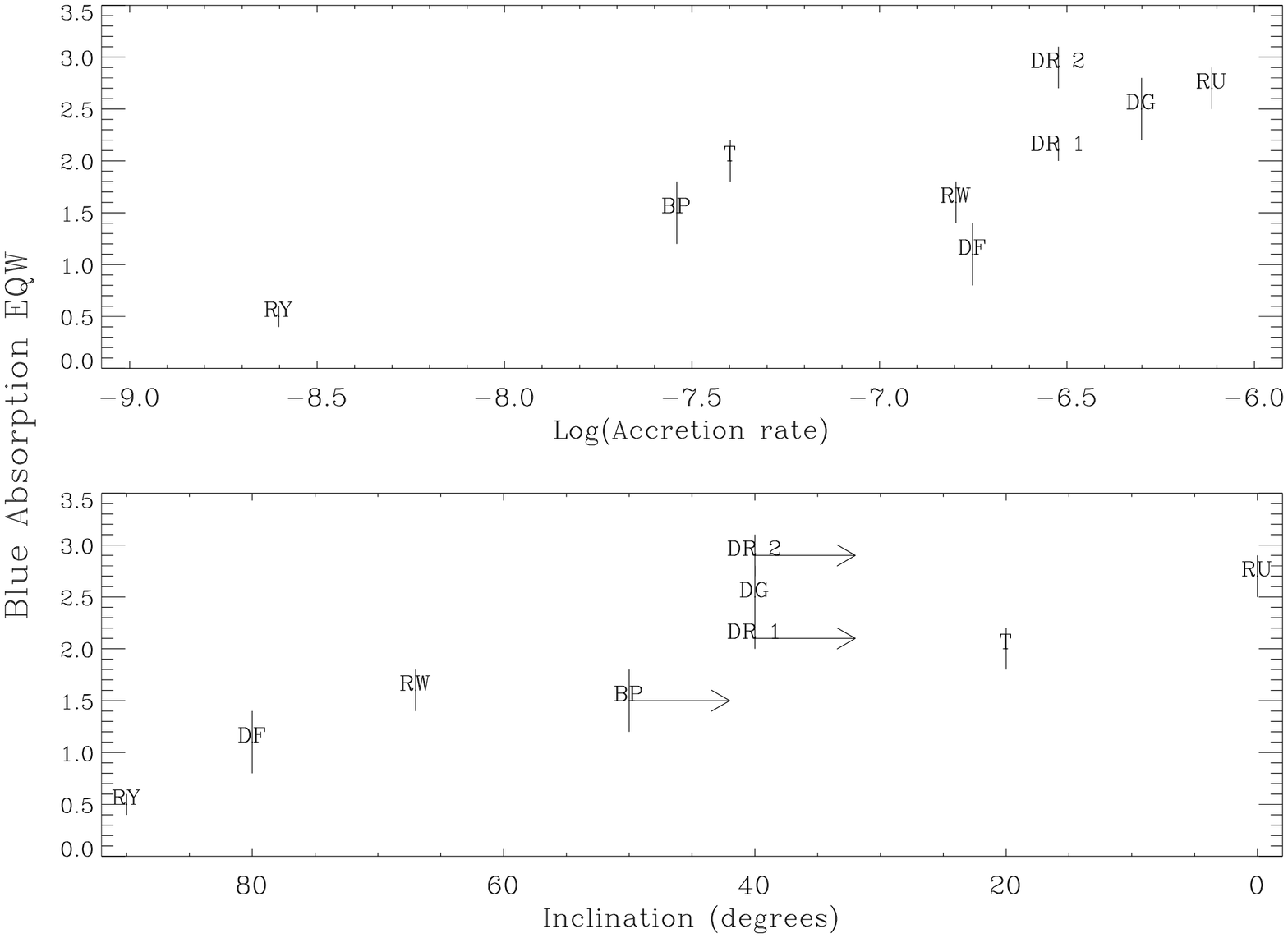}
\caption{\label{eqw_inc}EQW of wind absorption (in \AA) versus accretion rate (in $\msun / yr$, top) and stellar inclination (in degrees, bottom). The EQW has been corrected for the narrow central absorption in all stars. For the stars with double wind components, the EQW is taken over both. Upper limits in the inclination are indicated by arrows. Both epochs of DR Tau observations are indicated.}
\end{figure}

In Figure \ref{eqw_inc} (bottom) we compare the EQW of the blueshifted absorption with the stellar inclination. We have corrected for the narrow central absorption in all the stars. The figure indicates that edge-on stars have absorptions with less EQW than face-on stars. This is to be expected if the \mgii\ wind is collimated toward the poles. Note, however, the large scatter near 40 degrees. Part of it is likely due to differing accretion rates. The top plot of Figure \ref{eqw_inc} shows the relationship between accretion rate and wind EQW. Note that DG Tau has larger accretion than BP Tau, but both have similar inclinations. So, for a given inclination, larger accretion rate implies larger wind EQW. In a similar way, part of the scatter of the top plot of Figure \ref{eqw_inc} is due to the inclination: for a given accretion rate, two stars wind absorptions will have different EQW for different inclinations.

This simple picture is reinforced by the two observations of DR Tau. The second epoch observations shows indicators consistent with larger accretion rate: stronger \mgii\ line and larger continuum. So, as expected, the wind absorption has a larger EQW. Here we should remember that accretion rates vary (here they seem to vary by a factor of 2, as this is the factor necessary to make the second DR Tau observation fall in the correlation) and the adopted values are not necessarily valid for the epochs of our spectra.

All the stars with EQW$>2$ \AA\ have absorptions that reach zero. These tend to be low inclination, high accretion rate stars, as expected from a collimating wind that cools off at it expands. Note also that, in general, larger equivalent width implies larger terminal velocities: RY Tau has a wind terminal velocity of only -100 \kms\ while RU Lup has -350 \kms. In other words, the wind accelerates outward (as assumed before), and the maximum velocity component parallel to the disk is of the order of a hundred kilometers per second.

We have intentionally avoided any discussion of an specific wind model, in order to highlight model-independent results. Within the context of the X-wind model \citep{1994ApJ...429..781S} it is possible to obtain launching velocities of the order of hundreds of \kms\ and accelerations up to some hundreds of \kms. While this is a ``cold wind'' model, dust grains evaporated by X-ray emission may provide the necessary \mgii\ populations (Al Glassgold, personal communication). The X-wind is launched close to the star. In contrast, models like those described by \citet{1990RvMA....3..234C} are disk wind models, originating far from the central star. While kinematically it may be possible to produce the observed velocities, it is not clear what source of heating (perhaps magnetic dissipation?) would ionize Mg in quantities large enough to produce the observed absorptions.

\section{Conclusions}

In this paper we continue the analysis (the first part of which is in Paper I) of GHRS observations of TTSs. We have observed eight CTTSs (BP Tau, T Tau, DF Tau, DR Tau, DG Tau, RW Aur, RY Tau, RU Lup) and one WTTS (HBC 388). The GHRS data consists of a spectral ranges 40 \AA\ -wide centered on 2800 \AA. For four of the CTTS we have simultaneous optical observations which contain \hal, \hbeta, \hei, \nad, and the \caii\ infrared triplet. The strongest feature in the UV spectra is the \mgii\ resonance doublet.

The \mgii\ resonance doublet lines for the CTTSs are wide ($\sim200$ to $\sim300$ \kms) and symmetric lines, and all are optically thick. They show strong wind signatures, as expected given that \mgii\ is a resonance line. The lines for the WTTS are much narrower ($\sim70$ \kms) and with no wind signature. We conclude that most of the line in the CTTSs is due to accretion related processes. This is a tentative conclusion, given that the atmospheres of the WTTS (a K1 star) and those of the CTTSs are bound to be very different. The WTTS also shows evidence of chromospheric downdraft (infall), as observed in other WTTSs.

The \mgii\ lines show a narrow absorption feature superimposed to the wide emission line. This line is wider than anything that could be produced ISM clouds described in the literature. This leads us to conclude that the narrow absorption feature is due to non-LTE effects (i.e., central core self-reversals) and not to the ISM. Furthermore, the WTTS in the sample shows a narrow central absorption that is the narrowest of the sample, which implies that the absorption is affected by accretion processes. We do not observe any correlation between accretion and line parameters, however. This may indicate that the accretion process is fairly uniform in angle.  

The \mgii\ emission lines can be described by a single Gaussian emission component plus one or more absorption components. The components are optically thick and, for the stars with the largest accretion rates the k component of the doublet is narrower and blueshifted with respect to the h component. Given the small sample presented here, it is not possible to draw general conclusions from this behavior, but it should be explained by any model of the \mgii\ lines. No correlation is found between accretion rate and the flux in the \mgii\ lines: for this small sample, the lack of correlation is consistent with the MAM. We do not find any correlation between any parameter of the \mgii\ line emission and the inclination, which again points to the fact that the emission comes from a region that illuminates most of the sky.   

The lines from the \mgii\ resonance doublet tend to closely follow the shape of \hal. The fact that the correspondence between \hal\ and \mgii\ is so good cast doubts on the argument that the large widths of the \hal\ lines are due to the linear Stark effect, and supports the idea that supersonic turbulence is the main mechanism responsible for the line widths.

The \mgii\ lines sample larger volumes of wind than the Balmer lines, as evidenced by the large, deep wind absorptions. Terminal wind velocities from 100 to 400 \kms are observed. We observe multiple wind components in BP Tau and DG Tau, where two different blueshifted absorptions at large velocities are clearly separated. We observe evidence of thermalization in the \hal\ wind close to the star and variations in the wind density and/or terminal velocity in a period of two years (for DR Tau). By measuring the equivalent width of the wind absorption we argue that the different morphologies of the wind can be explained by accretion rate and inclination effects: in general, for two stars with the same inclination, the one with the larger accretion rate will have a wind signature with larger equivalent width. We believe that the correlation between equivalent width and inclination provides a simple, model-independent confirmation that the wind in CTTS is not spherical. 

In Paper I we showed that the \civ\ and \siiv\ lines seem to have multiple components in some stars. The parameters of these components are very different from those of the optical \caii\ and \hei\ lines, as expected given the very different thermal regimes they sample. However, models and observations of these optical lines have shown that their narrow components are likely to be formed close to the star, perhaps not unlike the \civ\ and \siiv\ lines. The disagreement in the kinematic parameters of the optical and UV lines remain a mystery. 

As this work shows, simultaneous comparison between wide ranges of lines in TTS can be very fruitful. We have shown here that while the magnetospheric accretion model is the canonical model used to understand TTS many observations remain to be explained. Higher resolution data and better models will be needed before we can say that we fully understand T Tauri stars. The current consensus, based on the magnetospheric accretion model, is that the emission components of the strong optical lines observed in TTS come only from the funnel flow, with the T Tauri wind providing absorption but not emission. Such simple picture has been motivated by the failure of (spherical and conical) emission wind models to reproduce observed profiles. However, no unified model (emission in the funnel, emission and absorption in the wind) has been attempted with a realistic, self-consistent wind model, like the X-wind model. Such next generation models are a necessary step in our understanding of these systems.

\acknowledgments

This paper is based on observations made with the NASA/ESA
{\it Hubble Space Telescope}, obtained at the Space Telescope Science Institute, which is operated by the Association of Universities for Research in
Astronomy, Inc., under NASA contract NAS 5-26555. DRA and GB acknowledge support from the National Science Foundation (grant ASI952872) and NASA (grant NAG5-3471). FMW acknowledges support through NASA grants NAG5-1862, STScI-GO-5317-0193A, STScI-GO-058750294A, and STScI-GO-3845.02-91A to SUNY Stony Brook. CMJK acknowledges support through NASA grant NAG5-8209.

We are very grateful to the Lick observers Geoff Marcy, Dan Popper (deceased), and Lawrence Aller for generously agreeing to observe our targets during their scheduled observing time. Celso Batalha has provided us with an spectrum of RU Lup in advance of publication. We have benefited from fruitful exchanges with Seth Redfield and Al Glassgold.

\bibliographystyle{apj}
%\bibliography{/garavito/ardila/tex/bibtex/ardila}

%\documentclass{aastex}
%\begin{document}

\begin{deluxetable}{lccccl} 
\tabletypesize{\small}
\tablecolumns{6} 
\tablewidth{0pc} 
\tablecaption{Log of Observations\label{tab_log_mg}}
\tablehead{
\colhead{Target (Date)} &\colhead{$\lambda_{cen}$} & 
   \colhead{UT start} & \colhead{Exposure} & 
   \colhead{Readouts} & \colhead{Root}\\
\colhead{} &\colhead{(\AA)} & 
   \colhead{(UT)} & \colhead{(sec.)} & 
   \colhead{} & \colhead{} }
\startdata 
\cutinhead{Pre-COSTAR}
RU Lup (8/24/92)&  2798.7  &  6:20:48 &  230 &  1 & Z10T010B \\
        & \\
BP Tau (7/30/93)&  2799.0  & 17:38:47 &  563 &  2 & Z18E0104 \\
        & \\
RW Aur (8/10/93)& 2799.1  & 22: 3:17 &  281  &  1 & Z18E0404\\
        & \\
DR Tau (8/5/93) &  2799.0 &  0:43: 5  &  281 &  1 & Z18E0304 \\
        & \\
DF Tau (8/8/93)&  2799.2  & 2:35:25  &  563 &  2 & Z18E0204 \\
        & \\
RY Tau (12/31/93) & 2798.9  &  9: 3: 6 &  230 &  1 & Z1E7010A \\
\cutinhead{Post-COSTAR}
DR Tau (9/7/95)  &  2799.3 & 17:35:49  & 1075 &  7 & Z2WB0204 \\
        &  2799.4 & 22:37:31  &  614 &  4 & Z2WB020B \\
        & \\
HBC388 (9/9/95)&  2799.5  & 19:19: 4 &  1024 &  5 & Z2WB0409\\
       & \\
T Tau (9/11/95)  &  2799.2  &  1:55:55 &  409 &  4 & Z2WB0304 \\
        & \\
DG Tau (2/8/96)&  2798.8  &  1:51:18 &  614  &  4 & Z2WB0104 \\
&  2799.0  & 11:48: 6 &  537  &  3 & Z2WB010D \\  %s-w=1.23px (4.7 km/s)

\enddata
\end{deluxetable}
%\end{document}
%\documentclass{aastex}
%\begin{document}
%\newcommand \siiv{\ion{Si}{4}~}
%\newcommand \civ{\ion{C}{4}~}

\begin{deluxetable}{lccc} 
\tablecolumns{4} 
\tablewidth{0pc} 
\tablecaption{\label{table_opt_log}Optical Observations} 
\tablehead{ 
\colhead{Name} & \colhead{Date-UT}   & \colhead{Exposure (sec.)} & \colhead{Observer}  }
\startdata 
BP Tau & 7/31/93 - Average of 11:30 and 11:46 & 1800 & Marcy, G. \\
DF Tau & 8/08/93 - 11:48  & 1200 & Aller, L. \\
RW Aur & 8/10/93 - 11:57  & 1800 &Marcy, G. \\
DR Tau & 8/04/93 - 11:46, 8/05/93 - 12:03& 1718, 1164 &Popper, D. \\
\enddata
\tablecomments{All observations were made with the Lick 3m Shane telescope, using the Hamilton echelle. DR Tau was observed on two consecutive days.}
\end{deluxetable} 

%\end{document}

%\documentclass{aastex}
%\begin{document}
%\newcommand \heI{\ion{He}{1}}
%\newcommand \hal{H$\alpha$}
%\newcommand \hbeta{H$\beta$}
%\newcommand \hgamn{H$\gamma$}
%\newcommand \hdeln{H$\delta$}
%\newcommand \kms{km s$^{-1}$}
%\newcommand \nosp{$\!\!\!\!$}
%\newcommand \htwo{H$_2$~}
%\newcommand \mgii{\ion{Mg}{2}~}

\begin{deluxetable}{ccccc} 
\tabletypesize{\scriptsize}
\tablecolumns{5} 
\tablewidth{0pc} 
\tablecaption{\label{all_lines}Line Measurements} 
\tablehead{\colhead{Line} & 
\colhead{Center\tablenotemark{a}}& 
\colhead{FWHM\tablenotemark{b}} &
\colhead{Flux \tablenotemark{c}}& 
\colhead{Unred. Flux \tablenotemark{c}}  }
\startdata 
\cutinhead{BP Tauri}

\mgii\ k (2795.53 \AA) &$-40\pm3$ &$256\pm3$ & $171\pm6$ & $430\pm20$  \\
Blue Abs. 1 &$-71.6\pm0.6$ & $20\pm2$ & $V_\infty=-350$ \kms \\
Blue Abs. 2 &$-129\pm2$ & $140\pm7$ &$V_\infty=-90$ \kms\\
\mgii\ h (2802.70 \AA) &$-50\pm4$&$262\pm3$ & $120\pm5$ & $300\pm10$  \\
Blue Abs. 1 &$-72.2\pm0.6$ & $15\pm2$ &$V_\infty=-350$\kms &\nodata\\
Blue Abs. 2 &$-134\pm2$ & $146\pm8$ & $V_\infty=-90$\kms  &\nodata\\
\hal (6562.80 \AA)&$-12\pm2$ & $251\pm6$ &\nodata &\nodata \\
\hbeta (4861.32 \AA)\tablenotemark{d}&$-15\pm4$& $170\pm10 $ &\nodata&\nodata  \\
HeI (5876 \AA) NC &$ -2\pm1 $&$44\pm4$ &\nodata&\nodata \\
BC &$4\pm13$      &$180\pm50$  &\nodata&\nodata \\
CaII (8498 \AA) NC &$0.7\pm0.6$&$22\pm2$&  \nodata &\nodata \\
 BC &$-10\pm5$ & $140\pm10 $& \nodata &   \nodata \\
\cutinhead{T Tauri}
\mgii\ k (2795.53 \AA) & $-3.8\pm0.5$ & $ 225.7\pm0.6 $ & $1006\pm9$&$21400\pm200$ \\
Blue Abs.&$V_\infty=-300$\kms & \nodata& \nodata& \nodata \\

\mgii\ h (2802.70 \AA) & $-11.9\pm0.5$ & $219.9\pm0.5$ & $810\pm7$ & $17000\pm200$\\
Blue Abs. & $V_\infty=-300$\kms &\nodata &  \nodata& \nodata \\

\cutinhead{DF Tauri}
\mgii\ k (2795.53 \AA) & $-22\pm1$ & $207\pm2$ & $85\pm3$ & $191\pm6$ \\
Blue Abs. & $-60.6\pm0.6$ & $98\pm2$& $V_\infty=-250 $\kms &  \nodata\\
\mgii\ h (2802.70 \AA) & $-27\pm2$ & $191\pm2$ & $71\pm3$ & $159\pm7$ \\
Blue Abs. & $-58.2\pm0.9$ & $104\pm4$ &$V_\infty=-250 $\kms & \nodata\\
\hal (6562.80 \AA) Emission & $-14.5\pm0.5$ & $220\pm1$ &\nodata& \nodata \\
Blue Abs. &$V_\infty=-120 $\kms&\nodata&\nodata&\nodata \\
\hbeta (4861.32 \AA) Emission &$-24\pm2 $ & $130\pm5 $ &\nodata&\nodata  \\
Blue Abs. &$V_\infty=-120 $\kms& &\nodata  &\nodata \\
HeI (5876 \AA) NC & $7\pm1$ & $28\pm4$ &\nodata&\nodata \\
              BC & $30\pm5$ & $55\pm5$ &\nodata&\nodata \\
CaII (8498 \AA) NC & $ 11\pm2 $ & $23\pm5$ &   \nodata&   \nodata \\
                BC & $14\pm5$ & $100\pm20$ & \nodata &   \nodata \\
CaII (8542 \AA) NC &$ 8\pm2$ & $45\pm6 $  & \nodata &   \nodata  \\
		BC & $20\pm10$ & $160\pm40 $ & \nodata &   \nodata  \\
CaII (8662 \AA) NC & $ 11\pm2$ & $ 24\pm5 $ & \nodata &   \nodata  \\ 
		BC & $ 14\pm5$ & $ 110\pm20 $ & \nodata   &   \nodata  \\ 

\cutinhead{RW Auriga}

\mgii\ k (2795.53 \AA)  & $-11.2\pm0.7$ & $373\pm2$ & $1100\pm20$ & $8500\pm200$ \\

Blue Abs. & $V_{\infty}=-210$\kms & \nodata &\nodata &\nodata \\
\mgii\ h (2802.70 \AA)  & $-3.7\pm0.6$ & $395\pm2$ & $1240\pm10$ & $9600\pm100$ \\

Blue Abs. & $V_{\infty}=-200$\kms & \nodata &\nodata &\nodata \\

\hal (6562.80 \AA)&$-1.2\pm0.4$&$334\pm2$&\nodata&\nodata \\
HeI (5876 \AA) NC &$1\pm6$  &$40\pm30$ &\nodata&\nodata \\
              BC & $-1\pm7$ & $100\pm200$  &\nodata&\nodata \\
Red absorption &$220\pm40$ & $30\pm90$ &   \nodata&   \nodata \\
CaII (8498 \AA) NC &$-134\pm3$  &$68\pm7$   &   \nodata&   \nodata \\
                BC &$18\pm2$  &$169\pm7$    & \nodata &   \nodata \\
\cutinhead{DG Tauri\tablenotemark{e}}

\mgii\ k (2795.53 \AA) &$-62.3\pm0.7$ & $261\pm1$ & $284\pm3$ & $5040\pm50$ \\
Blue Abs. 1 &$-36.5\pm0.5$ & $82\pm1$ & $V_\infty=-100$ \kms & \nodata \\
Blue Abs. 2 &$-106\pm1$ & $176\pm2$ &$V_\infty=-330$ \kms &\nodata \\
\mgii\ h (2802.70 \AA)  &$-21\pm1$ & $283\pm1$ & $155\pm2$ & $2720\pm40$ \\
Blue Abs. 1 &$-36.2\pm0.7$ & $70\pm2$ &$V_\infty=-100$ \kms& \nodata \\
Blue Abs. 2 &$-104\pm3$ & $151\pm5$ &$V_\infty=-350$ \kms & \nodata \\

\cutinhead{DR Tauri (8/5/93)}

\mgii\ k (2795.53 \AA)& $120\pm1$ & $178\pm3$ & $35\pm1$ & $298\pm8$ \\
Blue Abs. &$V_\infty=-350$ \kms& \nodata& \nodata& \nodata \\
\mgii\ h (2802.70 \AA) &$120\pm2$ & $168\pm4$ & $24\pm1$ & $205\pm7$ \\
Blue Abs.  &$V_\infty=-350$ \kms&\nodata& \nodata& \nodata \\

\hal (6562.80 \AA) NC & $121.5\pm0.1$ & $ 117.5\pm0.4 $ & \nodata & \nodata \\
BC & $ 50.1\pm 0.3$ & $470.5\pm0.4$ & \nodata & \nodata \\
Absorption & $-148.6\pm0.8$ & $150\pm2$ &\nodata &  \nodata \\
HeI (5876 \AA)NC &$6\pm1$& $41\pm3$ &\nodata &\nodata \\
              BC &$-18\pm7$&$170\pm20$ &\nodata &\nodata\\
Redshifted Abs. & $230\pm10$ & $100\pm30$ &\nodata &\nodata\\

\cutinhead{DR Tauri (9/7/95)\tablenotemark{e}}

\mgii\ k (2795.53 \AA)  &$115\pm1$ & $180\pm3$ & $12\pm1$ & $108\pm9$ \\
Blue Abs. &$V_\infty=-450$ \kms&\nodata& \nodata& \nodata\\ 
\mgii\ h (2802.70 \AA)  &$121\pm3$ & $167\pm3$ & $10\pm1$ & $86\pm8$ \\

Blue Abs. &$V_\infty=-450$ \kms&\nodata & \nodata& \nodata\\

\cutinhead{RY Tauri}

\mgii\ k (2795.53 \AA)  &$-0.4\pm0.9$ & $298\pm2$ & $158\pm3$ & $266\pm5$ \\
Absorption &$-11.1\pm0.7$ & $66\pm2$ & $V_\infty=-100$\kms & \nodata \\
\mgii\ h (2802.70 \AA)  &$-2\pm1$ & $297\pm3$ & $111\pm2$ & $187\pm3$ \\
Absorption & $-11.4\pm0.9$ & $62\pm3$ & $V_\infty=-100$\kms & \nodata \\

\cutinhead{RU Lupi}
\mgii\ k (2795.53 \AA) &$-26\pm2$ & $277\pm3$ & $1170\pm30$ & $11700\pm300$ \\
Blue Abs. & $V_\infty=-350$\kms  & \nodata  & \nodata  & \nodata \\
\mgii\ h (2802.70 \AA)  &$-11\pm2$ & $303\pm2$ & $980\pm20$ & $9700\pm200$ \\
Blue Abs. &$V_\infty=-350$\kms   & \nodata  & \nodata  & \nodata \\

\cutinhead{HBC 388}
\mgii\ k (2795.53 \AA) & $6.3\pm0.4$&$70.9\pm0.7$&$18.3\pm0.5$&$21.9\pm0.6 $\\
Red Abs.& $31.5\pm0.8$&$20\pm2$&\nodata&\nodata\\
\mgii\ h (2802.70 \AA) &$3.2\pm0.3$& $62.4\pm0.7$& $13.2\pm0.3$ & $15.8\pm0.4$ \\
Red Abs. & $29.5\pm0.8$& $16\pm2$&\nodata &\nodata \\
\enddata
\tablecomments{All wavelengths are are wavelengths in the stellar rest frame. The errors in each measurement are indicated. For \mgii\ h \& k the quoted parameters are those of the fitted Gaussian emission components only. The final flux does not include absorptions or the contribution of the UV 3 multiplet members. We use the reddenings quoted in Paper I}

\tablenotetext{a}{Units are \kms.}
\tablenotetext{b}{Units are \kms.}
\tablenotetext{c}{Units are $10^{-14}\rm{\ ergs\ sec^{-1}\ cm^{-2}}$.}
\tablenotetext{d}{Part of the profile falls outside the detector.}
\tablenotetext{e}{For this epoch, we have two spectra centered around 2800 \AA. The two have been averaged before the Gaussian fit.}
\end{deluxetable}
%\end{document}

%\documentclass{aastex}
%\begin{document}

\begin{deluxetable}{ccccccc}
\tabletypesize{\scriptsize}
\tabcolsep 1pt
\tablewidth{0pt}
%\newcommand \msun{M_\odot}
%\newcommand \rsun{R_\odot}
%\newcommand \kms{km s$^{-1}$}

%\tablenum{1}
\tablecaption{\label{table_ism}Characteristics of Narrow Central Absorption}
\tablehead{
\colhead{Name} & \colhead{v } & \colhead{FWHM} & \colhead{EQW} & \colhead{v$_{r}$} & \colhead{v$_{ISM}$} &\colhead{v$_{ISM}-$v$_r$ } \\
\colhead{} & \colhead{(\kms) } & \colhead{(\kms)} &\colhead{(\AA)}&\colhead{(\kms)}& \colhead{(\kms)} &\colhead{(\kms)}}   
\startdata
BP Tau\\
k & $7.4\pm0.3$ & $30\pm1$ & $0.17\pm0.03$ &15.8& 24.6 &8.8\\
h & $6.9\pm0.4$ & $24\pm1$ & $0.13\pm0.03$\\
T Tau \\
k & $9.8\pm0.2$ & $25.0\pm0.7$ &$0.10\pm0.02$&19.1 & 25.3& 6.2\\
h & $6.1\pm0.2$ & $15.0\pm0.6$ &$0.10\pm0.02$ \\
DF Tau\\
k &  $4.8\pm0.4$ & $26\pm1$& $0.15\pm0.05$&15.8 &25.0 & 9.2\\
h & $4.0\pm0.5$ & $24\pm2$& $0.12\pm0.05$& \\
RW Aur \\
k &  $7.6\pm0.5$ & $20\pm2$ &$0.15\pm0.02$ &16 & 24.8 & 8.8\\
h & $8.6\pm0.8$ & $17\pm2$ &$0.12\pm0.02$\\
DG Tau \\
k & $0.3\pm0.2$ & $21\pm1$ &$0.22\pm0.1$ & 15.9 &25.0 & 9.1\\
h & $-1.9\pm0.3$ &$18\pm1$ & $0.18\pm0.1$\\
DR Tau (8/5/93)\\
k & $0\pm2$ & $28\pm5$ &$0.18\pm0.04$&27.6& 25.6 &-2.0\\
h & $0\pm3$ & $31\pm6$ &$0.09\pm0.01$ \\
DR Tau (9/7/95)\\
k & $-3.1\pm0.7$ & $26\pm2$ &$0.31\pm0.01$&27.6&25.6&-2.0\\
h & $-3.7\pm0.8$ & $21\pm2$&$0.26\pm0.07$  \\

RY Tau\\
k &  $-12.4\pm0.6$ & $64\pm2$&$0.56\pm0.05$&16.4  &24.7 & 8.3\\
h & $-12.5\pm0.8$ & $61\pm3$&$0.46\pm0.05$ \\
RU Lup\\
k & $-9.5\pm0.3$ & $33.4\pm0.7$&$0.30\pm0.05$&-0.5 &-22.9,-26.4& -22.4,-25.9\tablenotemark{a} \\
h & $-9.0\pm0.3$ & $31.6\pm0.7$&$0.30\pm0.05$ \\
HBC 388\\
k &  $8.0\pm0.3$ & $18.3\pm0.7$ &$0.19\pm0.03$&15.4& 25.4 &10 \\
h & $7.3\pm0.2$ & $15.4\pm0.7$&$0.12\pm0.04$ \\
\enddata
\tablecomments{The ISM velocities are the velocities predicted by the Colorado Model of the LIC \citep{2000ApJ...534..825R}. v$_r$ is the radial velocity of the star, quoted in Paper I, Table 1.}
\tablenotetext{a}{The two velocities quoted are for the LIC and the G cloud respectively}

\end{deluxetable}
\tabletypesize{\normalsize}

%\end{document}

\end{document}